\documentclass[twoside,twocolumn,9pt]{article}
\usepackage{extsizes}
\usepackage[super,sort&compress,comma]{natbib} 
\usepackage[version=3]{mhchem}
\usepackage[left=1.5cm, right=1.5cm, top=1.785cm, bottom=2.0cm]{geometry}
\usepackage{balance}
\usepackage{times,mathptmx}
\usepackage{sectsty}
\usepackage{graphicx} 
\usepackage{lastpage}
\usepackage{amsfonts}
\usepackage[format=plain,justification=justified,singlelinecheck=false,font={stretch=1.125,small,sf},labelfont=bf,labelsep=space]{caption}
\usepackage{float}
\usepackage{fancyhdr}
\usepackage{fnpos}
\usepackage[english]{babel}
\usepackage[normalem]{ulem}

\addto{\captionsenglish}{%
  
}
\usepackage{array}
\usepackage{charter}
\usepackage[T1]{fontenc}
\usepackage[usenames,dvipsnames]{xcolor}
\usepackage{setspace}
\usepackage[compact]{titlesec}
\usepackage{hyperref}

\usepackage{epstopdf}

\definecolor{cream}{RGB}{222,217,201}

\begin{document}

\pagestyle{fancy}
\thispagestyle{plain}
\fancypagestyle{plain}{


\renewcommand{\headrulewidth}{0pt}
}

\makeFNbottom
\makeatletter
\renewcommand\LARGE{\@setfontsize\LARGE{15pt}{17}}
\renewcommand\Large{\@setfontsize\Large{12pt}{14}}
\renewcommand\large{\@setfontsize\large{10pt}{12}}
\renewcommand\footnotesize{\@setfontsize\footnotesize{7pt}{10}}
\makeatother

\renewcommand{\thefootnote}{\fnsymbol{footnote}}
\renewcommand\footnoterule{\vspace*{1pt}%
\color{cream}\hrule width 3.5in height 0.4pt \color{black}\vspace*{5pt}} 
\setcounter{secnumdepth}{5}

\makeatletter 
\renewcommand\@biblabel[1]{#1}            
\renewcommand\@makefntext[1]%
{\noindent\makebox[0pt][r]{\@thefnmark\,}#1}
\makeatother 
\renewcommand{\figurename}{\small{Fig.}~}
\sectionfont{\sffamily\Large}
\subsectionfont{\normalsize}
\subsubsectionfont{\bf}
\setstretch{1.125} 
\setlength{\skip\footins}{0.8cm}
\setlength{\footnotesep}{0.25cm}
\setlength{\jot}{10pt}
\titlespacing*{\section}{0pt}{4pt}{4pt}
\titlespacing*{\subsection}{0pt}{15pt}{1pt}

\newcommand{\be}{\begin{equation}}
\newcommand{\ee}{\end{equation}}
\newcommand{\ba}{\begin{eqnarray}}
\newcommand{\ea}{\end{eqnarray}}
\newcommand{\la}{\langle}
\newcommand{\ra}{\rangle}
\newcommand{\rr}{\mathbf{r}}
\newcommand{\qq}{\mathbf{q}}

\fancyfoot{}
\fancyfoot[LO,RE]{\vspace{-7.1pt}\includegraphics[height=9pt]{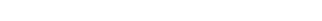}}
\fancyfoot[CO]{\vspace{-7.1pt}\hspace{13.2cm}\includegraphics{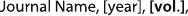}}
\fancyfoot[CE]{\vspace{-7.2pt}\hspace{-14.2cm}\includegraphics{head_foot/RF}}
\fancyfoot[RO]{\footnotesize{\sffamily{1--\pageref{LastPage} ~\textbar  \hspace{2pt}\thepage}}}
\fancyfoot[LE]{\footnotesize{\sffamily{\thepage~\textbar\hspace{3.45cm} 1--\pageref{LastPage}}}}
\fancyhead{}
\renewcommand{\headrulewidth}{0pt} 
\renewcommand{\footrulewidth}{0pt}
\setlength{\arrayrulewidth}{1pt}
\setlength{\columnsep}{6.5mm}
\setlength\bibsep{1pt}

\newcommand{\joe}[1]{\textcolor{blue} {#1}} 
\newcommand{\joec}[1]{\textcolor{blue} {[Joe: #1]}} 
\newcommand*{\cmb}[1]{{\color{BurntOrange}{#1}}}

\makeatletter 
\newlength{\figrulesep} 
\setlength{\figrulesep}{0.5\textfloatsep} 

\newcommand{\topfigrule}{\vspace*{-1pt}%
\noindent{\color{cream}\rule[-\figrulesep]{\columnwidth}{1.5pt}} }

\newcommand{\botfigrule}{\vspace*{-2pt}%
\noindent{\color{cream}\rule[\figrulesep]{\columnwidth}{1.5pt}} }

\newcommand{\dblfigrule}{\vspace*{-1pt}%
\noindent{\color{cream}\rule[-\figrulesep]{\textwidth}{1.5pt}} }

\makeatother

\twocolumn[
  \begin{@twocolumnfalse}
  

\begin{tabular}{m{4.5cm} p{13.5cm} }

& \noindent\LARGE{\textbf{Phenol release from
pNIPAM hydrogels: Scaling Molecular Dynamics simulations with Dynamical Density Functional Theory$^{\dagger}$}} \\
\vspace{0.3cm} & \vspace{0.3cm} \\

 & \noindent\large{H. A., P\'erez-Ram\'irez,$^{\ast}$$^{\textmd{a}}$ A. 
Moncho-Jord\'a,$^{\textmd{b}}$, G. 
Odriozola,$^{\textmd{a}}$$^{\ast}$ } \\

& \noindent\normalsize{We employed Molecular Dynamic simulations (MD) and Bennett's acceptance ratio method to compute the free energy of transfer, $\Delta G_{\rm{trans}}$, of phenol, methane, and $5$-Fluorouracil (5-FU), between bulk water and water-pNIPAM mixtures of different polymer volume fractions, $\phi_p$. For this purpose, we first calculate the solvation free energies in both media to obtain $\Delta G_{\rm{trans}}$. Phenol and $5$-FU (a medication used to treat cancer) attach to the pNIPAM surface so that they show negative values of $\Delta G_{\rm{trans}}$ irrespective of temperature (above or below the lower critical solution temperature of pNIPAM, $T_c$). Conversely, methane switches the $\Delta G_{\rm{trans}}$ sign when considering temperatures below (positive) and above (negative) $T_c$. In all cases, and contrasting with some theoretical predictions, $\Delta G_{\rm{trans}}$ maintains a linear behavior with the pNIPAM concentration up to large polymer densities. We have also employed MD to compute the diffusion coefficient, $D$, of phenol in water-pNIPAM mixtures as a function of $\phi_p$ in the diluted limit. Both $\Delta G_{\rm{trans}}$ and $D$ as a function of $\phi_p$ are needed inputs to obtain the release halftime of hollow pNIPAM microgels through Dynamic Density Functional Theory (DDFT). Our scaling strategy captures the experimental value of 2200 s for 50~$\mu$m radius microgels with no cavity, for $\phi_p \simeq$ 0.83 at 315 K.
}
\end{tabular}
 \end{@twocolumnfalse} \vspace{0.6cm}
  ]

\footnotetext[2]{Accepted for publication in \textit{Soft Matter}.
This is a preprint of an article published in \textit{Soft Matter}.
DOI: \href{https://doi.org/10.1039/D2SM01083F}{10.1039/D2SM01083F}.}


\renewcommand*\rmdefault{bch}\normalfont\upshape
\rmfamily
\section*{}
\vspace{-1cm}


\footnotetext{$^{a}$~Área de F\'isica de Procesos Irreversibles,
División de Ciencias B\'asicas e Ingenier\'ia, Universidad Aut\'onoma
Metropolitana-Azcapotzalco, Avenida San Pablo 180, 02200 Ciudad de M\'exico, 
Mexico. }
\footnotetext{\textit{$^{b}$~Departamento de F\'isica Aplicada, Universidad de 
Granada, Campus Fuentenueva S/N, 18071 Granada, Spain.}}
\footnotetext{\textit{$^{\ast}$email: godriozo@azc.uam.mx}}

%
%



\section{Introduction}
Hydrogel particles in their different morphologies (such as core-shell and hollow) have a wide variety of uses, e. g. drugs carrying and delivery~\cite{Kim1992,Qiu2001,Hoare2008,Kokardekar2012,Qian2013,Bae2013}, filters~\cite{Chu2001,Chu2004,Yu2017}, wastewater cleaning~\cite{Si2019}, cell culture~\cite{Cabral2011,Zhernenkov2015}, wound healing~\cite{Slaughter2009,Xu2020}, etc.~\cite{Mias2008,Koetting2015}. Depending on its purpose, the hydrogel is designed to load, release, or trap large cargo molecules. Drug delivery processes are key to achieving the desired therapeutic effect. Pharmacokinetics and target specificity can be controlled through the employment of drug carriers, such as liposomes, micelles, microspheres, and microgels. On the other hand, cargo uptake may occur directly from bulk towards hydrogel particles, or one can design gel membrane reactors able to concentrate the desired molecule to purify water. In particular, hydrogel particles made of Poly N-Isopropylacrylamide (pNIPAM) crosslinked polymer chains constitute a model thermoresponsive nanosystem. Indeed, pNIPAM is a type of temperature-sensitive polymer having a lower critical solution temperature in water, $T_c$, around $305$~K~\cite{Heskins1968,Longhi2004,Walter2010}. Below it, the polymer swells while hydrating and forming a translucent and homogeneous solution. Conversely, temperatures above $T_c$ make the pNIPAM-water mixtures phase separate, producing pNIPAM aggregates that disperse light and turn the previously translucent solution into a white milky dispersion~\cite{Heskins1968,Otake1990}. When pNIPAM is synthesized employing crosslinkers, it is possible to form core-shell and (following a somewhat more complicated procedure) hollow hydrogel particles that share the same swelling ability~\cite{Nayak2005,Qian2013,Contreras2015,Brugnoni2018}. The pNIPAM matrix of the hydrogel structure includes covalent bonds so that it remains well-defined even after following several reversible heating and cooling cycles around $T_c$~\cite{Otake1990,Zhang2003}.

In many of the applications mentioned above, controlling the kinetics of cargo load and release is essential to optimize the efficiency of these nanoparticles. This type of process may have characteristic times that range from the order of seconds or minutes while being controlled by diffusion and molecular interactions. Hence, it comes as no surprise that predicting the release kinetics of a given cargo molecule from its microparticle capsule is a complex task. In fact, the design of efficient nanocarriers usually entails extensive experimental work mostly based on trial and error~\cite{Li2016}. To help the development of rational designing of these nanosystems, we need reliable theoretical models able to predict the load/release kinetics as a function of the observable parameters.

There are two descriptors used in physical sciences and material engineering that determine the kinetics of cargo molecules through complex media, such as a polymer network of a hydrogel nanoparticle. The first one is the reversible work required to transfer a given molecule (cargo) from one bulk to another at constant temperature and pressure, known as the free energy of transfer, $\Delta G_{\rm{trans}}$. For an infinitely dilute cargo system, this free energy equals the effective interaction difference between the two bulks and the molecule, $u_{\rm{eff}}$, and dictates the partition coefficient (the concentration ratio between two bulks at thermodynamic equilibrium), a central quantity to design capsules and membranes with different materials and purposes~\cite{Teng2011,Kanduc2017,Adroher-Benitez2017b,Kim2017,Perez-Mas2018,Kanduc2019,MonchoJorda2019,MonchoJorda2020,Kanduc2021}. However, $\Delta G_{\rm{trans}}$ is a thermodynamic property and does not provide information about how long it takes to load or release the cargo from a given capsule or the mass flux across a membrane. Consequently, we also need a transport property, the diffusion coefficient, $D$, to access the load (or release) kinetics~\cite{Domb1990,Saltzman1991,Gehrke2006,Hansel2018,Kanduc2018}.

Therefore, a complete theory aimed to describe the load/release process requires a proper estimation of $\Delta G_\textrm{trans}$ and $D$ and then being able to consider these two parameters in the governing kinetic equations. These two properties strongly depend on the volume fraction occupied for the polymer chains, $\phi_p$. As already mentioned, $\phi_p$ is a function of temperature~\cite{Yang2017} due to the deswelling of the polymer network above $T_c$. However, for a fixed temperature, $\phi_p$ also changes when moving from the denser interval region of the hydrogel particle to its more diluted external shell, in which $\phi_p$ gradually tends to zero. In addition, $\phi_p$ can also be controlled by the number of crosslinkers used in the particle synthesis: a higher crosslinker concentration leads to hydrogel particles with a smaller swelling ratio and larger $\phi_p$~\cite{Lopez2019,Mashudi2021}.

In this work, we investigate these two properties and explore the load/release kinetics for cargo molecules diffusing across a pNIPAM (Poly N-Isopropylacrylamide) hydrogel particle. For this purpose, we use a combination of atomistic Molecular Dynamics simulations (MD) and Dynamical Density Functional Theory (DDFT). MD simulations allow us to determine directly $\Delta G_{\rm{trans}}$ and $D$. DDFT is a theory able to predict the time evolution of the non-equilibrium density profiles of the cargo concentration~\cite{Marconi1999}, which has been applied successfully to similar problems~\cite{Angioletti2014,AngiolettiUberti2018,MonchoJorda2019,MonchoJorda2020}. We use the values of $\Delta G_{\rm{trans}}$ and $D$ obtained from MD simulations as input parameters to obtain the time evolution of the cargo concentration. In particular, we calculate $\Delta G_{\rm{trans}}$ for several representative small cargo molecules, namely methane, phenol, and 5-fluorouracil (5-FU). On the one hand, methane is a small non-polar molecule, which changes its $\Delta G_{\rm{trans}}(\phi_p)$ sign when considering temperatures above and below $T_c$. On the other hand, phenol and 5-FU are ring polar molecules able to form hydrogen bonds with pNIPAM and water. Phenol is a water pollutant~\cite{Pan2008,Gong2015,Wang2022}, and microgels could assist water treatment processes. 5-FU is a drug commonly used to treat some kinds of cancer~\cite{Puga2013,Qi2017,Sabbagh2020,Dalei2020}. We fix two temperatures below and above $T_c$, given by $295$~K and $315$~K, and explore the dependence of the transfer free energy with $\phi_p$. We also calculate $D(\phi_p)$ for the particular case of phenol to directly compare the outcomes for a transport process obtained from simulations and DDFT. Finally, once DDFT is validated, it is used to make general predictions for the upload and release kinetics for hollow hydrogel nanoparticles.

\section{Simulations}
\label{section:exp}

As mentioned, the transfer free energy and diffusion coefficient of a cargo molecule are key parameters that control the kinetics of non-equilibrium diffusive uptake/release processes. In particular, both quantities are required as inputs for the DDFT method to yield release times. In addition, since we assume a continuous pNIPAM density profile for DDFT calculations, we need both quantities as a function of polymer density. For this purpose, we follow the simulation procedure proposed by Kanduč \textit{et al.}~\cite{Kanduc2018,Kanduc2019}, by which two equilibrated simulation boxes are built, one filled with pure water and the other with the target polymer density. These boxes are used as initial conditions for the calculations of $\Delta G_\textrm{trans}$ and $D$.

Simulations are based on an atomistic model, where the OPLS-AA force field is used for the polymer, and water is given by the SPCE model. This force field and its versions are frequently employed for pNIPAM systems~\cite{Jorgensen1996,Tamai1996,Walter2012,Lorbeer2014,Rodriguez-Ropero2015,Botan2016,Adroher-Benitez2017,Veinticinco2019,Veinticinco2019b,Garcia2019,Tavagnacco2021,Tavagnacco2022} and provide the best results for capturing the coil-to-globule transition when employed with the TIP4P/ice water model~\cite{Tavagnacco2022}. A particular version of the OPLS-AA~\cite{Siu2012} with the TIP4P/ice can capture the pNIPAM in water behavior at different temperatures and pressures~\cite{Tavagnacco2021}. However, the OPLS-AA combination with SPCE water is also common~\cite{Tamai1996,Adroher-Benitez2017,Kanduc2017,Kanduc2018}. We employed the GROMACS 2020 package for all simulations~\cite{Gromacs_GPU}. We set a timestep of $1$~fs and use the velocity rescale algorithm as a thermostat, with coupling time $\tau_t=0.1$~ps at two temperatures, $295$~K and $315$~K. The Berendsen barostat is set with a reference pressure of 1 bar and coupling constant $\tau_p=1.0$~ps. Lennard-Jones and the real-space electrostatic interactions are truncated with a cutoff of $1.2$~nm while using long-range corrections to energy and pressure. We set the particle mesh Ewald technique to deal with electrostatic interactions by considering a grid Fourier spacing of $0.12$~nm in all directions. Periodic boundary conditions are also applied in all directions to mimic a bulk system. Below $305$~K, the system is expected to form a homogeneous mixture, while it should aggregate displaying inhomogeneities above it. Nonetheless, all boxes have a well-defined overall polymer density.

\begin{figure*}[t!]
\centering
\centering \resizebox{1.0\textwidth}{!}{\includegraphics{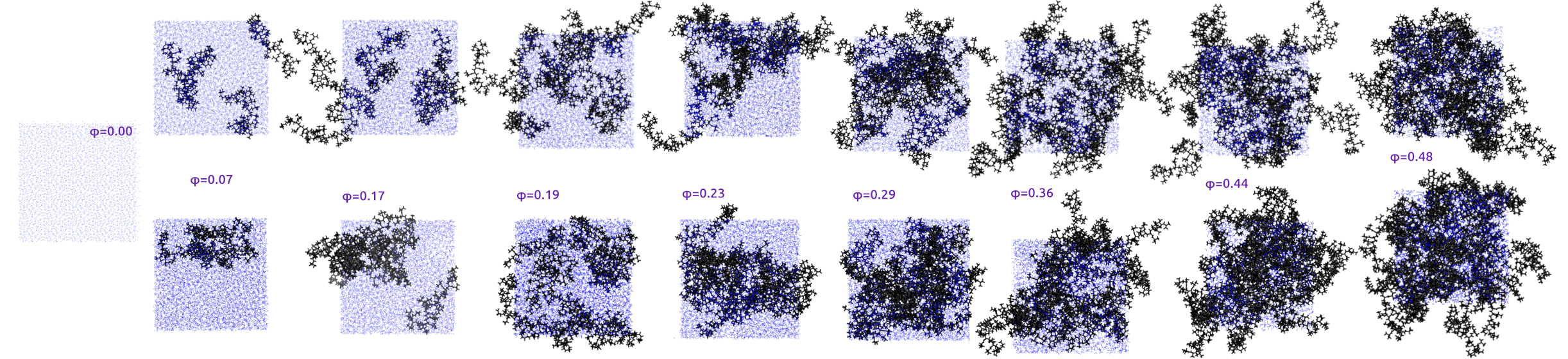}}
\caption{\label{fig1} Snapshots corresponding to the cells taken as the starting points for transfer free energy and diffusion coefficient calculations. The polymer density increases from left to right, as labeled. Temperature is 295 K for the top row and 315 K for the one at the bottom.}
\end{figure*}

To build the initial cells, we establish an empty cubic simulation box of $6$~nm per side, where we insert the necessary polymer atactic chains to approximately reach the target volume fraction. Each chain contains $20$ monomers. Note that chains are not crosslinked as implemented by Tavagnacco et al.~\cite{Tavagnacco2019}, which is a much more proper description of a microgel network. Crosslinks affect chain diffusivity by restricting reptation. In turn, cargo molecules attached to chains may increase their diffusion due to this fact. Simulation boxes are then filled with water and thermalized for $10$~ns in an NPT ensemble at $295$~K. At the end of this short thermalization, we check the polymer volume fraction and, if necessary, add more water or polymer. Then, each box is simulated for $100$~ns in an NPT ensemble to guarantee proper thermalization. Table~\ref{table1} summarizes the final amounts of oligomers and water molecules in each box. Fig.~\ref{fig1} shows a series of snapshots of every simulation box once thermalized at 295 K (top) and 315 K (bottom). Note that oligomers disperse at 295 K (below $T_c$) and aggregate at 315 K (above $T_c$). These cells serve as starting points for the following simulations.

\begin{table}[ht]
{\small
\begin{tabular}{||c c c||}
\hline 
Polymer chains  & Water molecules & $\phi_p$ \\ \hline
\hline
0  &   7000     & 0.00    \\ \hline
4  &   6550     & 0.07    \\ \hline
8  &   6037     & 0.17    \\ \hline
11 &   5560     & 0.19    \\ \hline
15 &   5225     & 0.23    \\ \hline
17 &   4600     & 0.29    \\ \hline
20 &   3865     & 0.36    \\ \hline
23 &   3156     & 0.44    \\ \hline
27 &   2313     & 0.48    \\ \hline
\end{tabular}
}
\caption{Polymer and water content in each box. The final and 
corresponding snapshots are depicted in Fig.~\ref{fig1}.}
\label{table1}
\end{table}

\subsection{Transfer free energy}
The solvation free energy is computed for different cargo molecules using Bennett's Acceptance Ratio method (BAR)~\cite{Bennett1976}. We calculate the solvation
free energy of cargo molecules for each polymer concentration, $\Delta G_{sol} 
(\phi_p)$. First, the simulation systems are prepared by 
randomly inserting cargo molecules in the previously built hydrogel boxes. For 
the smaller cargo molecules, we insert up to $15$ molecules to improve the sampling while maximizing their distances to mimic an infinite dilution. Due to this reason, we only insert $5$ to $8$ molecules for the largest cargo molecules.

Secondly, we define the initial state (A) as the one containing the cargo molecules and the final one as that without them (B). These two systems contain the same polymer volume fraction, $\phi_p$. The system change can be seen as a transformation in the total Hamiltonian, which switches off the cargo interaction with the other components (polymer and water molecules). In addition, we define a 
series of intermediate Hamiltonians, $H(\lambda)$, associated to a coupling parameter,
$\lambda$. The idea is to simulate several systems with different $\lambda$ 
values, which are then employed to estimate the free energy differences among them through
\begin{equation}
\Delta G_{\lambda_i} = G_{\lambda_{i+1}} - G_{\lambda_i} = \ln{\frac{Q_{NPT\lambda_i}}{Q_{NPT\lambda_{i+1}}}} ,
\end{equation}
where $Q_{NPT\lambda_i}$ and $Q_{NPT\lambda_{i+1}}$ are the isothermal-isobaric partition functions corresponding to the Hamiltonians defined for $\lambda_i$ and $\lambda_{i+1}$, respectively. The BAR method estimates $Q_{NPT\lambda_i}/Q_{NPT\lambda_{i+1}}$ as a ratio of weighted averages of the $\exp[-\beta (H(\lambda) + PV)]$ factors~\cite{Bennett1976}, where $\beta$ is the reciprocal temperature, $P$ is the pressure, and $V$ is the system volume. Note that the original method is deduced for the canonical ensemble instead of the isothermal-isobaric one~\cite{Bennett1976}. The sampling for these averages is carried out every 100 steps. $\lambda$ increases homogeneously in intervals of 0.0208 from 0 (A state) to 1 (B state), therefore we perform 48 simulations. We employ the \texttt{gmx bar} tool of Gromacs to yield $G_B-G_A=\sum_i \Delta G_{\lambda_i}$. In our case, the solvation free energy is given by $\Delta
G_{\rm{solv}}(\phi_p)=-(G_{B}-G_{A})$, given our selection of the initial and final states (we are passing the cargo molecules from bulk to vacuum), which coincides with the reversible work to move a cargo molecule from vacuum to bulk.

The coupling parameter $\lambda$ controls the Coulombic and Lennard-Jones 
interactions between the cargo molecule and the reference system. Coulombic and
Lennard-Jones interactions are decoupled separately through $48$ 
simulations, $24$ each, first Coulomb and then Lennard-Jones. Thus, the 
last simulation for Coulomb is the first for Lennad-Jones decoupling. The 
non-bonded interactions are smoothly decoupled, as $\lambda$ goes from $0$ to 
$1$. When the cargo molecules nearly disappear, soft-core interactions are
included to avoid numerical instabilities~\cite{Pham2012}. The alpha  parameter of the soft-core interactions, \texttt{sc-alpha}, is set to $0.5$ with a power \texttt{sc-power = 1} and a soft-core diameter of \texttt{sc-sigma=0.3}. This problem does not appear with Coulomb interactions given that Lennard-Jones are still turned on. The production simulations are $3$~ns long. Before that, we thermalize the system containing the inserted cargo molecules along 1000 ps in the NPT ensemble at the desired temperatures ($295$~K or $315$~K).

Note that for $\phi_p=0$, $\Delta G_{\rm{solv}}$ turns into the solvation free 
energy of cargo molecules in bulk water. The transfer free energy from a bulk of water towards another of polymer in water, at a certain $\phi_p$, 
is given by $\Delta G_{\rm{trans}}=\Delta G_{\rm{solv}}(\phi_p)-\Delta 
G_{\rm{solv}}(0)$. This quantity is the reversible work needed to move a given 
cargo molecule from water to the polymer-water mixture, and it is the 
parameter we need to feed the DDFT algorithm. It is worth mentioning that 
$\Delta G_{\rm{solv}}(0)$ must be precisely determined to achieve a low 
$\Delta G_{\rm{trans}}$ error. Thus, we repeat $15$ times the computation of 
$\Delta G_{\rm{solv}}(0)$ for each cargo and at least $5$ times for the other cases. From these repetitions (which go from the step of inserting the cargo molecules), we report the average and estimate the errors.

\subsection{Diffusion coefficient}
Aside from the effective interaction between polymer and cargo particles, we 
need to determine the diffusion coefficient of these molecules inside the 
hydrogel matrix. For this purpose, we calculate the mean squared displacement 
(MSD) of phenol in the hydrogel for different polymer volume fractions. 

As for free energy calculations, we use the prebuilt hydrogel boxes and randomly insert eight phenol molecules in each one. Systems are thermalized during $20$~ns in the NPT ensembles, at $295$~K or $315$~K, depending on the target temperature. Berendsen barostat and velocity rescale thermostat are set with coupling times of $0.1$ and $1$~ps, respectively. The production molecular dynamic simulation is performed along $800$~ns with a timestep of $1$~fs, on the NPT ensemble. To save disk space, only the coordinates of phenol are written every nanosecond during the simulation. The MSD of phenol is computed with the aid of the \texttt{gmx msd} tool of gromacs. The diffusion coefficient $D$ is calculated using the Einstein relation $\textrm{MSD}(t)=6Dt$, fitting a straight line to the long-time behavior of the $\textrm{MSD}(t)$.

\subsection{Diffusion through a pNIPAM membrane}
Finally, we design a simulation to directly compare its outcomes with the 
ones obtained from DDFT. In this case, we study the diffusion of phenol 
molecules from one side to the other of a slit-like membrane. The membrane is 
an array of polymer chains similar to those reported elsewhere~\cite{Adroher-Benitez2017,Veinticinco2019,Veinticinco2019b}. 
However, in this case, each polymer chain is made of $20$ monomers. Two layers of polymers are slightly overlapped along the $z$-axis, each layer being an array of $3 x 4$ chains. Water molecules are placed on both sides of the membrane and inside its interior. Impermeable walls are imposed at the borders $z=0$ and $z=l_z$, where $l_z$ is the length of the simulation cell. We have tried to minimize the adsorption of phenol molecules on them by imposing a soft repulsion. The cell has $l_x=l_y= 8.0$ nm and $l_z=20$ nm. In this case, we consider a larger number of reciprocal lattice vectors in the $z$-direction, $K_z$, so that $K_z=l_z K_x/l_x$. Periodic boundary conditions are applied only on the $x$ and $y$ axes. This way, the membrane-like array may represent a tiny region of the wall of a hollow and large hydrogel with its cavity separated from the exterior by a shell of polymer. We first insert all cargo molecules on the right side of the membrane (the outside of the hollow hydrogel). As the system evolves with time, cargo molecules should diffuse to reach the inner cavity of the membrane, passing through the membrane.

Previous to the insertion of cargo molecules, we performed a
semi-isotropic (1 bar and only $l_z$ is varied) thermalization of the 
water-membrane system during $50$~ns at $315$~K. We carried out a production run of $13$~ns with a time step one fs. From here, the density profiles of polymer, solvent, and cargo are computed as a function of time.

\section{Theory}
The dynamical density functional theory (DDFT) is used to trace the time 
evolution of the density profiles of the cargo in an out-of-equilibrium 
process. The mass accumulation at a given position $\textbf{r}$ and 
time $t$ is given by~\cite{Marconi1999,Angioletti2014,MonchoJorda2019} 
\begin{equation}\label{eq:dens-derivative}
 \frac{\partial \rho_c(\mathbf{r},t)}{\partial t} = -\nabla \cdot
\mathbf{J}_c(\mathbf{r},t),
\end{equation}
where $\mathbf{J}_c(\textbf{r},t)$ and $\rho_c(\mathbf{r},t)$ are the time and space-dependent mass flux and density of cargo molecules at $\textbf{r}$ and $t$, respectively. Assuming a Fickian process, the flux is proportional to the gradient of the chemical potential scaled by the inverse temperature
\begin{equation}\label{eq:flux}
 \mathbf{J}_c(\mathbf{r},t)= -D(\mathbf{r}) \rho_c(\mathbf{r},t) 
\nabla \beta \mu_c(\mathbf{r},t),
\end{equation}
where $D(\textbf{r})$ is the diffusion coefficient of 
the cargo molecules at position $\mathbf{r}$ and $\mu_c(\mathbf{r},t)$ is the chemical potential. Eqs.~\ref{eq:dens-derivative} and \ref{eq:flux} are key components of the DDFT method.

The other important component of the DDFT approach is related to finding an expression
for the chemical potential of the cargo, required to solve eqs.~\ref{eq:dens-derivative} and \ref{eq:flux}. The local chemical potential at time $t$ is given by
\begin{equation}\label{eq:chemical-potential}
 \mu_c(\mathbf{r},t) = \frac{\partial \mathfrak{F} 
[\rho_c(\mathbf{r},t)]}{\partial \rho_c(\mathbf{r},t)},
\end{equation}
where $\mathfrak{F}$ is a functional of $\rho_c(\mathbf{r},t)$. The main approximation of the DDFT method relies on the assumption that the equilibrium excess free energy functional can also be used in a non-equilibrium situation, replacing the equilibrium density profiles with non-equilibrium ones. Hence, the characteristic times of all varying external fields, such as the polymer profile, must be much larger than the diffusion times of the cargo molecules inside the gel (in our case, the external field, governed by the polymer profile, is fixed).

We consider three contributions to the free-energy-functional. The 
equation reads
\begin{multline}\label{eq:energy-functional}
 \beta \mathfrak{F} [\rho_c(\mathbf{r})] = \int\rho_c(\mathbf{r}) 
[\ln(\rho_c(\mathbf{r})\Lambda_c^3)-1] d\mathbf{r} + \\
 \int \rho_c (\mathbf{r})\beta u_{\rm{eff}}(\mathbf{r}) d\mathbf{r} + \int 
\beta 
f_{\rm{HS}}(\mathbf{r}) d\mathbf{r},
\end{multline}
where $\Lambda_c$ is the thermal wavelength of the cargo. The first term is 
the ideal gas contribution. The second term considers the interaction of the 
cargo particles with the polymer matrix, which can be seen as a fixed external 
potential. Finally, the last term accounts for the excess free energy due to 
the excluded volume of the other cargo particles. We are taking $f_{\rm{HS}}$ 
as given by the Carnahan-Starling free energy density to consider the excluded volume of the cargo molecule. Without this term, which is small at low cargo concentrations, the DDFT is equivalent to the Smoluchowski approach~\cite{Kim2022}.

Performing the functional differentiation, the corresponding chemical potential reads as~\cite{Angioletti2014,MonchoJorda2019,MonchoJorda2020}
\begin{eqnarray}
\label{muc}
\beta \mu_c(\mathbf{r},t) =\ln \left(
\rho_c(\mathbf{r},t)\Lambda_c^3\right)+\beta u_\textmd{eff}(\mathbf{r}) \nonumber
\\ + \phi_c(\mathbf{r},t)\frac{8-9\phi_c(\mathbf{r},t)+3\phi_c(\mathbf{r},t)^2}{
	(1-\phi_c(\mathbf{r},t))^3} ,
\end{eqnarray}
where $\phi_c=4\pi a_c^3 \rho_\textmd{c}/3$ and $a_c=0.25$ nm is the cargo radius.

The model reliability relies upon the accurate definition of the free energy 
functional. Since we focus on the absorption (desorption) of neutral molecules 
into (out of) neutral hydrogels, electrostatic interactions are neglected. In 
addition, the second term in eq.~\ref{eq:energy-functional} arises due to
the effective interaction potential between cargo and polymer. Thus,  
$u_{\rm{eff}}(\mathbf{r})$ is the necessary work needed to transport a cargo 
molecule from bulk water to position $(\mathbf{r})$ in terms of $k_BT$. This 
energy might be positive or negative since it results from the interplay 
between repulsive and attractive contributions to the overall force, which 
depends upon the chemical nature of the polymer, cargo, and solvent molecules. 
However, no further assumptions are needed for this term, as we obtain it 
directly from simulations (with a different name), $u_{\rm{eff}}(\mathbf{r})= \Delta G_{\rm{trans}}/N_A$, $N_A$ being the Avogadro's constant.

In addition to $\mu_c(\mathbf{r},t)$, eq.~\ref{eq:flux}
needs $D(\mathbf{r})$ as an input. As explained, we directly obtain this
parameter from molecular dynamics simulations through the MSD of the cargo 
molecules inside the hydrogel. More precisely, we obtain $D(\phi_p)$, while we
set $\phi_p(\textbf{r})$, as explained in the following subsection.    

We implement the DDFT algorithm in a homemade \texttt{c++} code, which makes 
use of CUDA to take advantage of Graphics Processing Units (GPUs). To this end, we discretize eqs.~\ref{eq:dens-derivative} and \ref{eq:flux} in space and time. Space is
discretized by employing grids of different mesh sizes, $\Delta$, the larger the gradients, the smaller $\Delta$. Units are rescaled for distances with $l_0=$ 1 nm, and for time with $\tau_0=l_0^2/D_0$, where $D_0$ is the cargo diffusion coefficient in bulk water, as computed from simulations. At the hydrogel interfaces, where gradients are large, the mesh size is minimum, $\Delta= 0.02l_0$. Conversely, inside and outside the gel, the mesh size can be larger. The time step is set in the interval 10$^{-4} \tau_0 <\Delta \tau<$ 
4$^{-4}\tau_0$. To improve efficiency, the idea is to set the maximum time step leading to smoothly varying (in time and space) cargo profiles.

\subsection{The hydrogel morphology}

Hydrogels can show several morphologies such as capsules, brushesgrafted on surfaces, hollow particles, membranes, etc~\cite{Roy2013,Koetting2015}. Here, we focus on the cases of hollow spherical hydrogels and planar membranes. The cores of the hollow hydrogels are just a cavity filled with solvent molecules. For spherical hydrogel particles, we set a spherical
coordinate system, whereas we employed a rectangular one for a membrane. For the former case, eq.~\ref{eq:dens-derivative} is rewritten as
\begin{equation}
 \frac{\partial \rho_c(r,t)}{\partial t} =-\frac{1}{r^2} 
\frac{\partial}{\partial r} \left(r^2 J(r,t)\right),
\end{equation}
where $r$ is the position measured from the center of coordinates placed at the 
hydrogel's center. For the second case, we have 
\begin{equation}
\frac{\partial \rho_c(z,t)}{\partial t} = -\frac{\partial 
J(z,t)}{\partial z}, 
\end{equation}
where $z$ is measured from the leftmost boundary of the simulation cell, to 
compare the outcomes with membrane simulations.  

We assume that the internal region of the hydrogel (and the membrane) have a uniformly 
distributed polymer volume fraction $\phi_p^{\rm{in}}$ and that this
concentration continuously vanishes towards the inside and the outside of 
the particle (membrane) throughout the corresponding internal and external hydrogel interfaces. Mathematically, such a profile 
can be given by  
\begin{equation}\label{hydrogelmorph}
\phi_p(x) = \phi_p^{\rm{in}} \frac{1}{2}[\rm{erf}(2(x-R_1)/ 
\delta)-\rm{erf}(2(x-R_2)/ \delta)],
\end{equation}
where $x=r$ ($x=z$) is the distance measured from the center of the hydrogel (from the
leftmost boundary of the simulation cell). $2\delta$ provides the thickness of the polymer interface (both inner and outer), 
and $R_1$ and $R_2$ define the position where $\phi_p(x) \approx 
\phi_p^{\rm{in}}/2$. In addition, $R_{\rm{in}}=R_1 - \delta$ is the largest $x$ 
value inside the cavity producing a $\phi_p(x)$ close to zero, and 
$R_{\rm{ext}}=R_2+ \delta$ can be interpreted as the hydrogel external radius 
(see the scheme inserted in Fig.~\ref{fig7}). This functional dependence for the polymer volume fraction represents a fair description of the core-shell morphology of the hydrogel particles, having a more concentrated, highly
cross-linked core, with uniform polymer mass distribution, surrounded by two fuzzy interfaces where the polymer density decays gradually to zero~\cite{Meyer2005,Berndt2006}.

\subsection{Cargo uptake}
Besides $D(\phi_p(x))$ and $\mu_c(\phi_p(x))$, we need 
boundary and initial conditions to solve eqs.~\ref{eq:dens-derivative} and \ref{eq:flux}.
For membrane and hydrogel particles, we are setting a null flux at $x=0$, 
\begin{equation}
J_c(x=0, t) = 0 \quad  \forall  t.
\end{equation}
Depending on the initial conditions, we can study cargo uptake or 
release. In the case of cargo uptake, all cargo particles are placed outside 
the hydrogel (or at the right of the membrane), and the cargo density profile 
is given by
\begin{equation}
\rho_c (x)= \left\{                                
\begin{array}{ll}
0 & x < R_{\rm{ext}} \\
\rho_c^{\rm{bulk}}  & x > R_{\rm{ext}}
\end{array}
\right. .
\end{equation}

For a diluted hydrogel suspension, far away from the hydrogel, the cargo 
concentration is constant and equal to the bulk concentration. We consider 
$x=5 R_{\rm{ext}}$ a far enough distance from the center of the hydrogel and 
fix it as the bulk concentration,
\begin{equation}
 \rho_c (x \rightarrow \infty,t)=\rho_c^{\rm{bulk}}.
\end{equation}
This condition implies no mass conservation and the establishment of a variable
mass flux from the outside. The flux from the outside turns zero only when 
reaching equilibrium. This situation is different from the membrane since our 
MD simulations are performed in the (anisotropic) NPT ensemble, meaning that 
the number of particles is kept constant and no mass flux flows into the 
system. Hence, in this case we set $J_c(x=L, t) = 0$,  $\forall  t$.

\subsection{Cargo release}
For desorption, we kept the first boundary condition; there 
is no flux at the center of the encapsulating particle (or at the leftmost 
boundary of the simulation cell for the membrane case). On the other hand, the 
cargo is initially distributed homogeneously through all the hydrogel, and 
therefore
\begin{equation}
\rho_c (x)=\left\{                                
\begin{array}{ll}
0 & x >  R_2 \\
\rho_c^{\rm{in}}  & x <  R_2
\end{array}
\right. .
\end{equation}
Again, we consider a diluted hydrogel suspension so that the cargo
concentration away from the central particle is practically zero, 
\begin{equation}
\rho_c (x \rightarrow \infty,t)=0.
\end{equation}
In this case, the mass flux flows from the inside to the outside of the system.

\section{Results}

\subsection{Solvation and transfer free energies}

\begin{figure}[t!]
\centering
\includegraphics[width=0.4\textwidth]{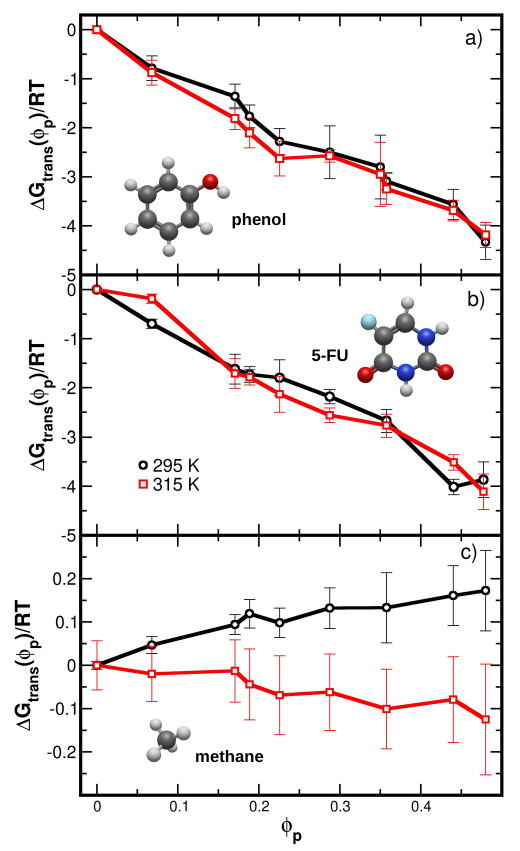}
\caption{\label{fig2} Transfer free energy from bulk water to a water polymer mixture for phenol, a), 5-FU, b), and methane, c), as a function of the polymer volume fraction, $\phi_p$. Black and red curves and symbols correspond to 295~K and 315~K, respectively. }
\end{figure}

We start presenting the MD simulation results for the transfer free energy at two temperatures, $T=$295~K and $T=$315~K, which correspond to a swollen and collapsed pNIPAM hydrogel particle, respectively. First, we show the solvation free energy in pure water and then include the pNIPAM chains for increasing values of $\phi_p$.

At 295~K, we have obtained the following values for the free energy of
solvation in pure water, $\Delta G_{\rm{solv}}=$ $(-20.41 \pm 0.04)$ kJ 
mol$^{-1}$, $(-107.9 \pm 0.1)$ kJ mol$^{-1}$, and $(9.30 \pm 0.03)$ kJ mol$^{-1}$, for phenol, 5-FU, and methane, respectively. The values can be compared with the experimental ones obtained at 298 K of $(-23 \pm 3)$ kJ mol$^{-1}$~\cite{Duarte2017}, $(-71 \pm 4)$ kJ mol$^{-1}$~\cite{Geballe2010}, and $(8.4 \pm 0.8)$ kJ mol$^{-1}$~\cite{Abraham1990}, respectively. In addition, the obtained values at $315$~K are $\Delta G_{\rm{solv}}=$ $(-18.42 \pm 0.06)$ kJ
mol$^{-1}$, $(-104.7 \pm 0.1)$ kJ mol$^{-1}$, and $(10.34 \pm 0.12)$ kJ mol$^{-1}$ for the same molecules. As can be seen, irrespective of temperature, phenol and 5-FU show negative values of $\Delta G_{\rm{solv}}$, whereas methane produces positive ones, a fact related to the different chemical nature of the molecules. Methane only interacts with water through van der Waals, contributing to reducing enthalpy while perturbing the high-energy hydrogen bond network of water molecules. However, water molecules somehow manage to avoid this hydrogen network  perturbation by reorganizing around methane~\cite{Graziano2019}. The net balance is a slight enthalpy decrease, $\Delta H_{\rm{solv}} < 0$. In addition, entropy is disfavored when a methane molecule is embedded into bulk water simply because it decreases the water available volume, which reduces its configurational entropy~\cite{Graziano2019,Graziano2000,Pica2016}, $\Delta S_{\rm{solv}} < 0$. This last term is common for all solutes since it is proportional to its solvent-accessible area. Hence, $\Delta G_{\rm{solv}}$ is governed by entropy for weakly interacting molecules with water, such as non-polar molecules, but turn enthalpy controlled for polar ones. Assuming $\Delta H_{\rm{solv}}$ and $\Delta S_{\rm{solv}}$ remain constant with temperature (they do not), we get $\Delta H_{\rm{solv}} =(-6 \pm 2)$ kJ mol$^{-1}$ and $\Delta S_{\rm{solv}} = (-0.052 \pm 0.008)$ kJ mol$^{-1}$ K$^{-1}$ from $\Delta G_{\rm{solv}}=\Delta H_{\rm{solv}}-T\Delta
S_{\rm{solv}}$, which reasonably compares with the experimental values of
$\Delta H_{\rm{solv}} =-10.9$ kJ mol$^{-1}$ and $\Delta S_{\rm{solv}} = -0.063$ kJ mol$^{-1}$ K$^{-1}$ at $298$~K~\cite{BenNaim1984}.
                 
Phenol and 5-FU are polar molecules, but more importantly, they can form 
hydrogen bonds with water acting both as acceptors and donors. Again, assuming
constant $\Delta H_{\rm{solv}}$ and $\Delta S_{\rm{solv}}$ values in the 295 - 
315 K range, we get $\Delta H_{\rm{solv}}= (-50 \pm 2)$ kJ mol$^{-1}$ and 
$\Delta S_{\rm{solv}} = -0.10 \pm 0.01$ kJ mol$^{-1}$ K$^{-1}$ for phenol. 
Thus, the absolute value of $\Delta S$ practically doubles as compared to 
methane, following its increased solvent-accessible area (from $1.39$~nm$^2$ to $2.58$~nm$^2$), and $\Delta H_{\rm{solv}}$ changes sign and increases an order of magnitude due to hydrogen bonding. Assuming hydrogen bonds with 20 kJ mol$^{-1}$ and neglecting other contributions to the enthalpy, phenol would produce an average of $2.5$ bonds with water, which roughly agrees with the $2.2$ bonds we have estimated with the {\it gmx hbond} tool. On the other hand, 5-FU, which has two oxygen, two nitrogen, and one fluorine, all hydrogen-bond forming atoms~\cite{Rosenberg2012}, leads to $\Delta H_{\rm{solv}}= (-155 \pm 4)$ kJ mol$^{-1}$ and $\Delta S_{\rm{solv}} = (-0.16 \pm 0.01)$ kJ mol$^{-1}$ K$^{-1}$. The $\Delta H_{\rm{solv}}$ would be consistent with the formation of around $7.7$ hydrogen bonds (again, under several assumptions) that is not that far from the $5.5$ ones we obtain with {\it gmx hbond}, which does not consider those linked to the fluor atom. In addition, it should be noted that the absolute value of $\Delta S_{\rm{solv}}$ is larger for 5-FU than for phenol, despite having similar solvent-accessible areas ($2.58$~nm$^2$ and 2.66 nm$^2$ for phenol and 5-FU, respectively). We believe that 5-FU further restricts the orientations of the first shell of water molecules than phenol through hydrogen-bond formation, which, in turn, explains this difference.

The $\Delta G_{\rm{trans}}$ values for the transfer of phenol, 5-FU, and methane from water to a water-polymer mixture with volume fraction $\phi_p$
are given in Fig.~\ref{fig2}. Phenol, 5-FU, and methane results are 
depicted from top to bottom. In each panel, the black and red curves 
correspond to $295$~K and $315$~K, respectively. A negative $\Delta G_{\rm{trans}}$ value means the molecule prefers the polymer mixture above water. Hence, phenol and 5-FU are more soluble when the polymer is present, for temperatures below and above the critical, $T_c \approx 305$ K. In addition, this preference does not seem to vary with temperature (we practically obtain the same curves irrespective of $T$, which means that the pNIPAM conformation, coil or globule, does not affect $\Delta G_{\rm{trans}}$). Note that the Gibbs-Helmholtz equation, $\partial (\Delta G_{\rm trans}/T)/\partial T |_P=-\Delta H_{\rm trans}/T^2$, applied to the transfer of phenol and 5-FU molecules yields a null $\Delta H_{\rm trans}$ value, an athermic process, and so $\Delta G_{\rm trans}=-T\Delta S_{\rm trans}$ for all $\phi_p$. Thus, the entropy increase is ruling the phenol and 5-FU adsorption in pNIPAM. A relatively large cargo molecule, as compared to water, increases the number of water micro-states when attaching to the pNIPAM surface. In addition, since $\Delta G_{\rm{trans}}$ controls the partition coefficient between the phases, this quantity stays constant when crossing $T_c$. Hence, there is no such thing as a squeezing effect~\cite{Gutowska1997}, at least for these particular molecules. Finally, $\Delta G_{\rm{trans}}(\phi_p)$ is linear for the whole $\phi_p$ we have studied. Indeed, we have conducted an extra determination for $\phi_p=0.75$ and phenol, which confirm the linear trend. This result is consistent with a virial expansion of the transfer free energy up to the first order of the polymer density, $ \Delta G_\textrm{trans}/RT \approx 2B_2 \rho_p$, where $\rho_p$ is the number density of monomers and $B_2$ is the second virial coefficient for the monomer-cargo effective interaction~\cite{Kim2020}.

It is worth mentioning that there exist theoretical models and coarse-grained computer simulation results predicting a non-monotonic behavior of $\Delta G_{\rm{trans}}(\phi_p)$ of molecules
with an effective attraction with the hydrated polymer matrix, showing a minimum at some intermediate polymer volume fraction, of about $\phi_p=0.1-0.3$ (which corresponds to a maximum in the partition coefficient)~\cite{Perez-Mas2018,Kim2020}. This minimum is followed by an increase of $\Delta G_{\rm{trans}}(\phi_p)$ with increasing $\phi_p$ due to volume exclusion effects induced by the presence of the polymer, which dominate over the attraction at high densities. However, these coarse-grained descriptions do not explicitly account for solvent molecules that also come into play in the free energy balance. In addition, we also think that the flexible polymer chains create available volume on demand by arranging and expelling water in the process.

Note that the curve corresponding to $\Delta G_{\rm{trans}}$ of phenol is very similar to the one obtained for 5-FU, which may seem surprising given the difference in $\Delta G_{\rm{solv}}$ we have found for bulk water. Indeed, as $\Delta G_{\rm{trans}}$ is the free energy difference between water-polymer and water solvation, the lesser number of hydrogen bonds formed when replacing water with pNIPAM should, in principle, disfavor 5-FU. However, the ring structure of 5-FU can overcompensate the loss of hydrogen bonds by providing a low interaction energy with pNIPAM, explaining the vanishing $\Delta H_{\rm{trans}}$ values. Since phenol and 5-FU show null $\Delta H_{\rm{trans}}$ values, it is not surprising that the corresponding $\Delta G_{\rm{trans}}=-T\Delta S_{\rm{trans}}$ yield similar values. $\Delta S_{\rm{trans}}$ depends on the size of the molecules, and they are similar.

\begin{figure}[t!]
\centering
\includegraphics[width=0.45\textwidth]{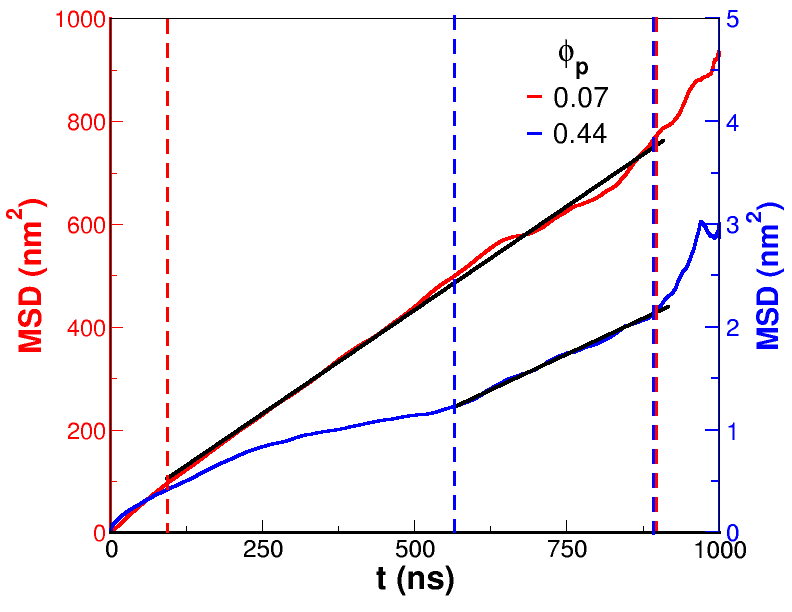}
\caption{\label{fig3} Mean square displacement (MSD) as a function of time, 
$t$, for diluted phenol in pNIPAM-water mixtures at 295 K. The red and blue 
curves correspond to $\phi_p=$ 0.07 and 0.44, as labeled. Note that the MSD 
scales differ a couple of orders of magnitude (red left, blue 
right).  }
\end{figure}

The case of methane is especially interesting, as $\Delta G_{\rm{trans}}$ changes sign with temperature, implying a weakly endothermic transfer. Of course, we attribute this change to the pNIPAM-water restructuring but not to methane itself. As mentioned, pNIPAM shows a conformation change that shifts from coil-to-globule when increasing temperature across $T_c$. When coil, pNIPAM is surrounded by water molecules so that methane cannot practically experience the difference between the bulk of water and the hydrated pNIPAM. Indeed, $\Delta G_{\rm{trans}}$ values are always tiny. The small positive $\Delta G_{\rm{trans}}$ values may only arise because detaching a water molecule from pNIPAM to make room is even slightly worse than making room inside pure water. But this situation changes (not very dramatically but enough to invert the sign) when $T>T_c$. In this case, the pNIPAM water mixture turns heterogeneous, where places of high and low local $\phi_p$ coexist. Then, if for instance, the molecule is transferred from bulk water to a high $\phi_p$ region, it suddenly observes a less-polar environment (a not-so-bad place to be from a non-polar perspective). Conversely, there is still a chance to transfer a methanol molecule from bulk water to a pNIPAM-water interface, which produces a positive $\Delta G_{\rm{trans}}$, but the pNIPAM solvent-accessible area is smaller at high $T$. The final possible moves of a methane molecule between similar regions yield no change. The average of the Gibbs changes associated with these possibilities weighted by their chances leads to a slightly negative $\Delta G_{\rm{trans}}$ showing a negative slope with $\phi_p$, as depicted by the red curve of the bottom panel of Fig.~\ref{fig2}.

\subsection{Diffusion coefficients}

\begin{figure}[t!]
\centering
\includegraphics[width=0.45\textwidth]{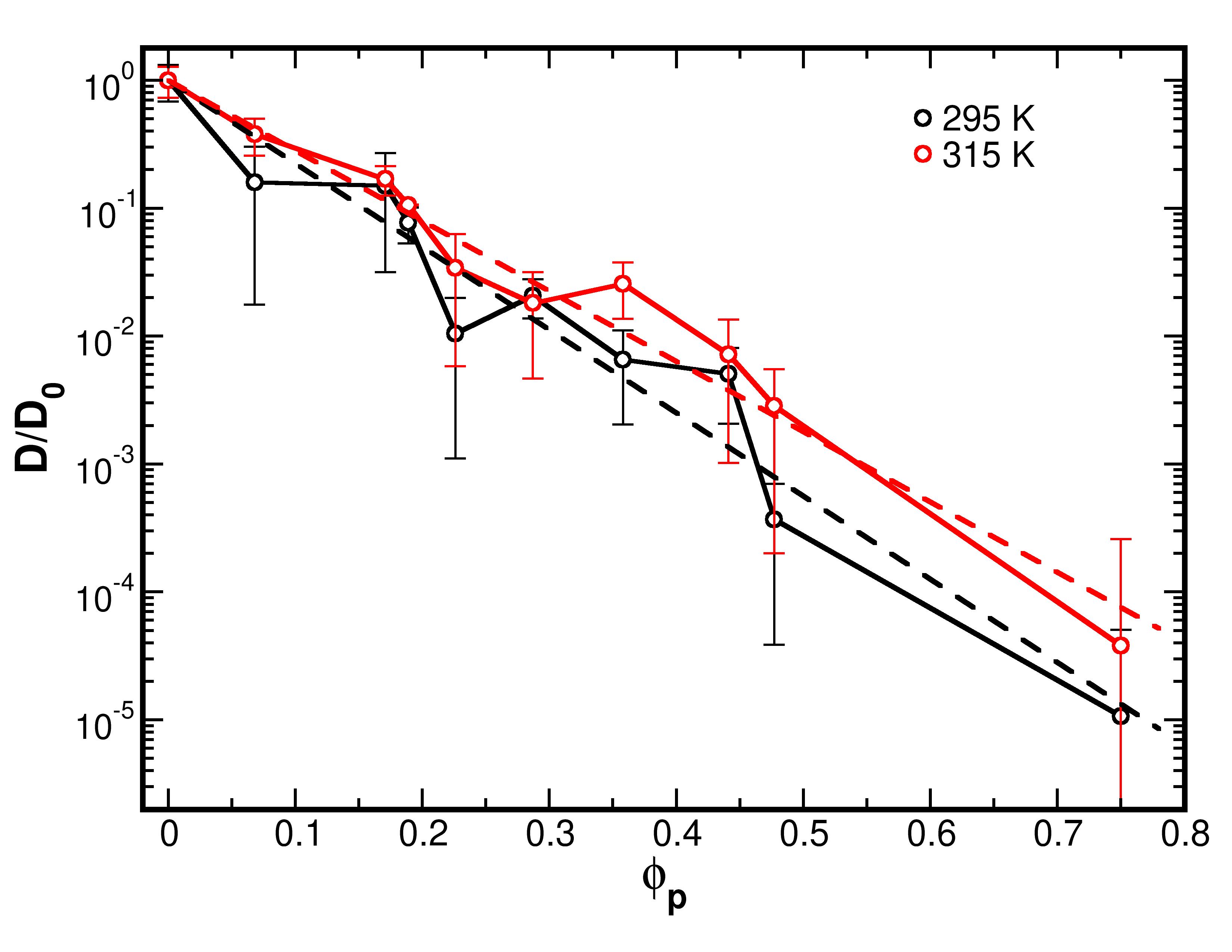}
\caption{\label{fig4} Normalized diffusion coefficients as a function of $\phi_p$ for $295$~K (black) and $315$~K (red). The black and red dashed lines are fits to the corresponding data, $D/D_0=10^{-6.5\phi_p}$ and $D/D_0=10^{-5.5\phi_p}$, respectively.    }
\end{figure}
       
In this Section, we present the simulation results regarding the calculation of the effective diffusion coefficient of the phenol inside the polymer matrix of the pNIPAM hydrogel, $D$. Fig.~\ref{fig3} depicts the mean squared displacement (MSD) of phenol in hydrated polymer bulk with different volume fractions, $\phi_p$, at $295$~K. For low $\phi_p$ values, the time behavior of the MSD looks like the red curve  of Fig.~\ref{fig3}, i.e. it shows a linear growth from the beginning up to long times. In such cases, it is easy to perform a linear fit to determine the effective diffusion coefficient of the cargo molecule, $D$.

Conversely, at 
high $\phi_p$ values, the MSD looks like the blue curve of Fig.~\ref{fig3}, 
where three stages appear. This sort of behavior has been previously observed in dense polymer networks by Kanduč \textit{et al.}~\cite{Kanduc2018}. They found that, under conditions of large polymer crowding, the water molecules distribute heterogeneously in fractal-like clusters embedded in the nanometer-sized cavities of the polymer matrix. The nanoclustered water dominates the diffusion of the cargo molecule, which occurs via a hopping mechanism through wet transition states: the cargo molecules hop from one cavity to another using transient water channels opened by polymer fluctuations. Indeed, in our computer simulation, we find a linear behavior of the MSD with $t$, which last around $100$~ns. In this time window, the molecules follow a diffusive regime before getting stuck in the polymer network. As mentioned in the previous section, phenol has a preference for the hydrated pNIPAM surface, replacing water molecules. When phenol attaches to the amide group of pNIPAM through hydrogen bonding, it stays there for a while trapped inside a cavity, leading to a continuous decrease of the MSD slope. At long times, phenol can hop from one cavity to another, regaining the diffusive (linear) regime. The long-time behavior defines $D$ from a linear fit of the MSD against $t$, as shown in Fig.~\ref{fig3}.

The obtained values of $D$ for phenol in water are $(0.96 \pm 0.35) \times
10^{-9}$ m$^2$ s$^{-1}$ and $(1.52 \pm 0.56) \times 10^{-9}$ m$^2$ s$^{-1}$ at 295 K and 315 K, respectively, which compare well with the experimental values of $(1.013 \pm 0.012) \times 10^{-9}$ m$^2$ s$^{-1}$ and $(1.802 \pm 0.036)
\times 10^{-9}$ m$^2$ s$^{-1}$ obtained at $298.65$~K and $323.5$~K~\cite{Winkelmann2018}. It is 
custom to normalize the diffusion coefficients with a reference value, $D_0$. In our case, we use the values obtained for pure water as a reference, each one at the corresponding temperature. 

In general, the diffusion of molecules inside hydrogels is a complex problem that involves several effects such as the hydrogel free volume reduction, enhancing the hydrodynamic drag of the molecule, the increased path length due to obstruction, and a combination of hydrodynamic drag and obstruction effects, among others~\cite{Amsden1998}. In addition, the relative importance of these effects depends on the polymer volume fraction. For small $\phi_p$, $D$ is expected to be controlled by the interaction of the cargo molecule with a single polymer chain. In contrast, for large $\phi_p$, the hopping mechanism turns dominant~\cite{Kanduc2021}. The dependence of $D$ on the polymer volume fraction is different for each of these mechanisms. In our case, the normalized curves $D/D_0$ for $295$~K and $315$~K as a function of $\phi_p$ follow an exponential decay law with $\phi_p$ (see Fig.~\ref{fig4}). The obtained exponential dependence interpolates smoothly between the low and high polymer density regime. The calculations performed at large $\phi_p$ required extending the simulations to very long times ($800$~ns) to capture several of these rare hopping events.

\begin{figure*}[t!]
\centering
\includegraphics[width=0.8\textwidth]{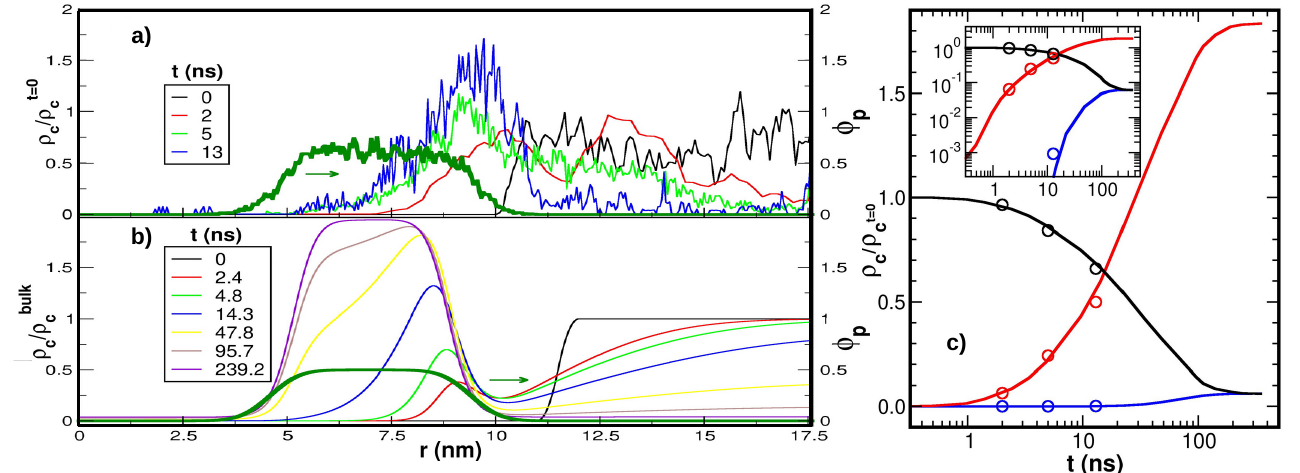}
\caption{\label{fig5} Normalized density profiles as a function of
time for the diffusion of phenol through a pNIPAM membrane at 315 K by means of MD, a), and DDFT, b). We use dark green and thick lines for the pNIPAM profile at time zero, which is set as similar as possible for MD and DDFT. The thin lines with different colors correspond to phenol profiles at different times, as labeled.  c) The normalized extent of membrane crossing. Black, red, and blue lines and symbols correspond to the right, membrane, and left cargo normalized densities. The normalization is made by taking the initial density at the right as a reference. Lines correspond to theory and symbols to simulations. The inset of panel c) shows the same data in a log-log representation.  }
\end{figure*}

As mentioned, both curves at $295$~K and $315$~K are normalized, so thermal agitation is already taken into account. Thus, differences between curves should be attributed to other factors. In addition, we cannot attribute the differences between both curves in terms of the effective attraction of phenol with pNIPAM with temperature, as we have observed practically the same transfer free energies in Fig.~\ref{fig2}. Hence, we need something else to explain why the curve corresponding to $315$~K goes above the one at $295$~K. Again, we ascribe this difference in behavior to the fact that pNIPAM chains collapse and aggregate above $305$~K, leading to large polymer and water-rich regions. Thus, $D$ measures a weighted average of molecules embedded in bulk water surrounded by polymer interfaces and those placed in polymer-rich regions. The firsts strongly contribute to $D$ and overcompensate the others. Conversely, at $295$~K, the chains are hydrated showing a coil conformation and well dispersed, defining a single homogeneous phase. Thus, polymer chains, the attractive obstacles that a phenol must overcome to diffuse, are everywhere, making them attach continuously to the much larger solvent-accessible surface area they display. This thermal behavior of $D$ is also observed for highly-grafted cylinders~\cite{Peng1998,Chu2001}.

\subsection{Comparing DDFT and MD outcomes}

Before scaling the MD results with DDFT, we should compare their results at least for a small out-of-equilibrium system. For this purpose, we have built a simulation prism with a pNIPAM membrane splitting space into two regions, one (at the right) having phenol distributed homogeneously with a 0.3 M concentration. The pNIPAM membrane has an internal polymer volume fraction of $\phi_p = 0.5$, which may represent a condition of collapsed microgels ($T=$ 315 K). During the simulations, we average the profile during $0.5$~ns to avoid noise while trying to make a short-time averaging. In addition, we compute the cargo  density integrals at the right, along, and left of the membrane and divide the results with the corresponding integral widths. We are taking the integral limits where $\phi_p = 0.25$ and then normalizing the results with the initial outside density. These computations are shown in panel c). The MD run lasted $13$~ns which is far from the time needed to achieve the equilibrium state. We can only observe the first stages, where phenol accumulates at the pNIPAM water interface and slightly diffuse through the membrane (see Fig.~\ref{fig5} a) and b)). At this stage, only a few molecules manage to cross it (see panel c)). To correctly compare the simulations with the DDFT predictions, the diffusion coefficient and effective cargo-polymer interactions that enter the DDFT differential equations are taken from the simulation data for $D(\phi_p)$ and $\Delta G_\textrm{trans}(\phi_p)$, respectively. The comparison with the DDFT outcomes is good. At $13$~ns, we have the same adsorption peak at the interface, which has ever increased its height from the initial state. Both peak heights also coincide during the time evolution of the cargo density profiles. In addition, the normalized extent of membrane crossing, as measured by the three curves of panel c), also shows similar results. Thus, we conclude that the general behavior is well-captured with our DDFT implementation. Note that we should have waited over $200$~ns to reach equilibrium, something we cannot afford with this system size.

The good quantitative agreement between DDFT and MD makes us gain confidence in eq.~\ref{eq:energy-functional}, despite the simplicity of the free-energy functional and that our theoretical model does not take into account an effective attraction between phenol molecules in water (which does exist, as they phase separate at large phenol concentrations). Both assumptions seem not to affect the results.

\subsection{Scaling MD through DDFT}

\begin{figure}[t!]
\centering
\includegraphics[width=0.5\textwidth]{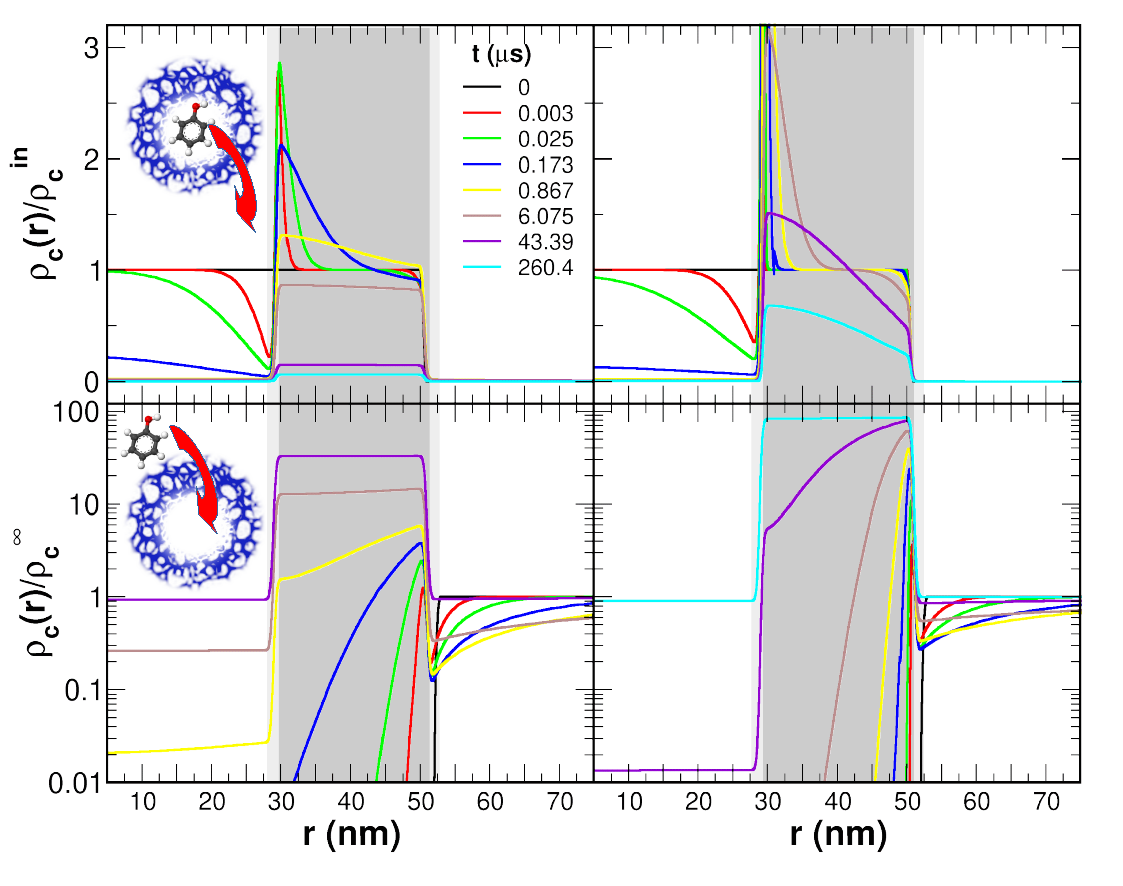}
\caption{\label{fig6} Normalized density profiles of phenol during 
release (top panels) and load (bottom panels) of a hollow microgel particle. 
The gray areas represent the walls of the microgel and the light gray the
interfaces. Lines of different colors correspond to the phenol profiles at 
different times. Left panels correspond to 295 K and right ones to 315 K. Note that bottom panels employ logarithmic scales for density profiles.}
\end{figure}

We proceed to study the behavior of larger systems. Let us consider a spherical and hollow nanogel with $R_{\rm{in}}=30$ nm and $R_{\rm{ext}}=50$ nm, at 295 K and 315 K. As the pNIPAM expands at $T<T_c$, we assume $\phi_p^{in}=$0.25 for 295 K and $\phi_p^{in}=$0.5 for 315 K, which in turn impact the values of $\Delta G_{\rm{trans}}$ and $D$. The actual values of $\phi_p^{in}$ of a real nanogel depends on the details of its synthesis, but there is typically more than a factor of two between expanded and shrunk networks.

Black lines Fig.~\ref{fig6} correspond to initial conditions. For release processes (see the upper panels), we assumed a homogeneous distribution of phenol inside the particle, including its walls. As in the previous section, we assumed $\phi_c^{t=0}=0.3$~M for both load and release processes. In the first stages of the release process, phenol depletes close to the internal interface on the cavity side, and a huge peak develops at the wall side. This happens due to the high effective attraction existing between phenol and pNIPAM. The peak is highest at around 0.025~$\mu$s, and then diffusion inside the wall makes it broaden and reduce its height. Meanwhile, some of the phenol mass goes outside the particle through the outer interface. As time goes by, the cavity load reduces its content, always showing a depleted region close to the inner interface that settles the gradient and flux direction, keeping a relatively high phenol concentration inside the pNIPAM matrix and a constant one at the external interface. Hence, while some phenol remains inside the cavity, the release rate is kept practically constant given that the outside concentration is always close to zero and the outer interface has a nearly constant concentration. When the cavity runs out of phenol, the phenol concentration of the wall starts decreasing, at around 6~$\mu$s, which, in turn, slows the release rate. This final stage takes a lot of time compared to the first stage.

The release process is strongly affected by temperature. As can be seen in the top-right panel, $260$~ns are enough to empty the microgel content at $295$~K but not at $315$~K (note that we assume a constant particle size). The longer release times occurring at high temperatures are due to the increase of the effective phenol-wall attraction and the decrease of the phenol diffusion coefficient inside the walls. Indeed, the diffusion coefficient shows a large impact at $315$~K since the phenol concentration is always decreasing at the outer interface despite the high peak at the inner one.        

The load process occurs by fixing a constant phenol concentration far away from the particle, an initial condition where the exterior has a constant $\phi_c^{t=0}=0.3$, and no phenol inside. Thus, the gradient and the flux are reversed. The external interface starts building a phenol peak, which spreads towards the inside of the wall. However, the peak never stops rising, and the spreading continuously increases the phenol concentration within the polymer network of the hollow hydrogel. Note that the phenol density at the wall reaches very high values (note the logarithmic scale). Meanwhile, some phenol gets inside the cavity, but most of it concentrates on the walls. When reaching equilibrium, the cavity and the exterior attain the same concentration, and most of the load locates at the walls. Thus, for pNIPAM particles in water, the load capacity of phenol is much higher when there is no cavity. Finally, the load process is faster at $295$~K ($44$~ns) than at $315$~K ($260$~ns) for the same reasons as the release. Note also the much larger final phenol concentration at the polymer matrix at 315 K than at $295$~K. These final uptake equilibrium states could also be considered as initial conditions for the release process.

\begin{figure}[t!]
\centering
\includegraphics[width=0.45\textwidth]{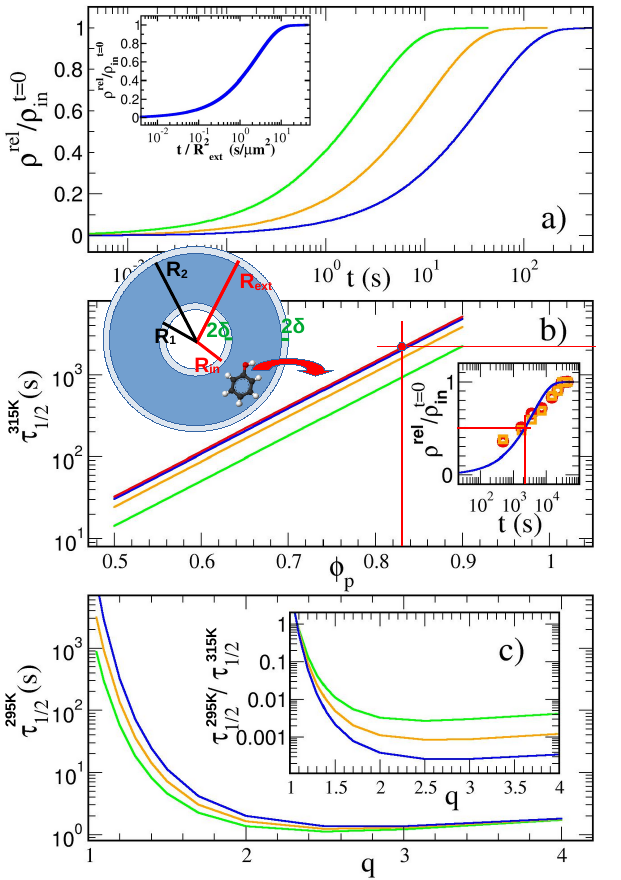}
\caption{\label{fig7} a) Normalized extent of release as a function of time from particles with $R_{\rm ext}=$ 1.0~$\mu$m (green line), 2.0~$\mu$m (orange line), and 4.0~$\mu$m (blue line) at 315 K and $\phi_p=0.75$. The inset shows the scaling of the curves by replacing $t$ with $t/R_{\rm{ext}}^2$ (the green and orange curves are below the blue one). b) Release halftime, $\tau_{1/2}$, for particles with $R_{\rm ext}=23 \mu$m and $R_{\rm{in}}/R_{\rm{ext}} =$ 0 (black line), 0.25 (blue line), 0.5 (orange line), and 0.75 (green line), at 315 K as a function of $\phi_p$. The red bullet corresponds to the experimental value reported by Si et al.~\cite{Si2019} as obtained at 318 K (we  neglected the temperature difference). The inset compares the theoretical extent of release for $\phi_p^{315\rm{K}}=0.83$ with the one corresponding to the T-MIH (red circles) and T-NIH (orange squares) particles~\cite{Si2019}. c) Release halftime, $\tau_{1/2}^{295\rm{K}}$, at 295 K for microgels without cavity, $R_{\rm ext}=23 \mu$m, and $\phi_p^{315\rm{K}}=$ 0.7 (green line), 0.8 (orange line), and 0.9 (blue line), as a function of $q$. The inset shows the release halftime ratio $\tau_{1/2}^{295\rm{K}}/\tau_{1/2}^{315\rm{K}}$ between swollen and collapsed states as a function of $q$. The drawing on top of b) is a scheme showing the different parameters of eq.~\ref{hydrogelmorph}.  }
\end{figure}

During a release process, one can integrate the density profiles inside the particle to calculate the load as a function of time. The normalized extent of release can then be computed as the difference between the initial load and the load at time $t$, divided by the former. This quantity is shown in Fig.~\ref{fig7} a) as a function of time for particles with no cavity, $\phi_p=$0.75, and different $R_{\rm{ext}}$ values. In a semi-log representation, all curves display a sigmoidal shape, signaling that the last release stage is very slow. All curves collapse by replacing $t$ with $t/R_{\rm{ext}}^2$ since the characteristic time is given by $\tau_0=l_0^2/D_0$ (inset of Fig.~\ref{fig7} a)). Hence, rescaling the particle size while keeping $R_{\rm{in}}/R_{\rm{ext}}$ fixed simply leads to a redefinition of the time step (neglecting the role played by $\delta$). For the same reason, the release halftime, $\tau_{1/2}$, also scales with $R_{\rm{ext}}^2$ and the density profiles of Fig.~\ref{fig6} remain the same when replacing $r$ with $r/R_{\rm{ext}}$ and $t$ with $t/\tau_0$. Indeed, the swelling kinetics of microgels also scales with $R_{\rm{ext}}^2$ since it involves the diffusion of small molecules through the particle~\cite{Tanaka1979,Wahrmund2009}. Thus, there is no need to solve qs.~\ref{eq:dens-derivative} and \ref{eq:flux} for micro-sized particles. For instance, from Fig.~\ref{fig7} a) we get $\tau_{1/2} = 1.462$s for $R_{\rm{ext}}=$1 $\mu$m, and thus a particle with $R_{\rm{ext}}=$23~$\mu$m has $\tau_{1/2} =$ 773~s for the release of phenol at 315 K and $\phi_p =$0.75, a smaller time than that experimentally reported at 318 K (around 2200 s)~\cite{Si2019}.

Taking the collapsed state at 315 K as a reference and defining $q$ as the degree of swelling of the microgel, calculated as the ratio of the radii at $T$ and 315 K, one gets $q=1$ at 315 K and $q>1$ at temperatures below the critical. For $q=1$, $R_2+\delta=$ 23 $\mu$m, $\delta=$ 10 nm, $D(\phi_p)=D_0^{315\rm{K}} 10^{-5.5 \phi_p}$ with $D_0^{315\rm{K}}=1.52 \times 10^{-9}$ m$^2$s$^{-1}$ (the fitted expression to the 315 K data of Fig.~\ref{fig4}), $u_{\rm eff}(\phi_p)= -9.0\phi_p$ (the fitted expression to the data of Fig.~\ref{fig2} a)), and $R_1/R_2=$ 0, 0.25, 0.5, and 0.75, we get from eq.~\ref{meanpassagetime2} (see the Appendix) the black, blue, orange, and green lines shown in Fig.~\ref{fig7} b) as a function of $\phi_p$, respectively. This panel shows that the cavity size plays a role. Increasing $R_{\rm{in}}/R_{\rm{ext}}$ leads to shorter release times because decreasing the wall thickness reduces the region where phenol has slow diffusivity due to its interaction with the polymer. More importantly, from this panel, we note that for a microgel without a cavity we need $\phi_p \simeq$0.83 to agree with the experimental release halftime reported by Si et al.~\cite{Si2019}. Although 0.83 looks large as compared to the experimentally reported values for the collapsed state around 0.7 of van Durme et al.~\cite{vanDurme2004} and Sasaki et al.~\cite{Sasaki1999}, Dong and Hoffman report 0.77~\cite{Dong1990} and computer simulation results are in the 0.76-0.94 range~\cite{Kanduc2018}. In addition, Si et al.~\cite{Si2019} report a thermo-gravimetric analysis of their T-MIH and T-NIH particles, where only 13$\%$ and 12$\%$ water weight is lost below 550 K, respectively. The comparison of the theoretical extent of release with the experimental data is given as an inset of Fig.~\ref{fig7} b) (neglecting the small temperature difference).

Finally, it is interesting to compute the release halftime for the collapsed state at 295 K  as a function of $q$, which is shown in Fig.~\ref{fig7} c). For this purpose, we set $D(\phi_p)=D_0^{\rm{295K}} 10^{-6.5 \phi_p}$ with $D_0^{295\rm{K}}=0.96 \times 10^{-9}$ m$^2$s$^{-1}$ (the fitted expression to the 295 K data of Fig.~\ref{fig4}), and $\phi_p^{295\rm{K}}=\phi_p^{315\rm{K}} q^{-3}$ with $\phi_p^{315\rm{K}}=$ 0.7 (green line), 0.8 (orange line), and 0.9 (blue line), and plotted $\tau_{1/2}^{295\rm{K}}$ as a function of $q$. Here, we have kept $R_2+\delta=$ 23 $\mu$m, $\delta=$ 10 nm, and function $u_{\rm eff}(\phi_p)$ fixed, as it does not practically change with $T$ (see Fig.~\ref{fig2} a)). Note that in our case, eq.~\ref{meanpassagetime2} simplifies since the second term turns dominant, as the first and third terms vanish for $R_1 \rightarrow \delta \rightarrow 0$, which allows us to write a simple expression for $\tau_{1/2}^{295\rm{K}}/\tau_{1/2}^{315\rm{K}} (q) \simeq q^2 D_0^{315\rm{K}} 10^{\phi_p^{315\rm{K}}(6.5/q^3-5.5)}/D_0^{295\rm{K}}$. We show this ratio as an inset of panel~\ref{fig7} c). As observed, $\tau_{1/2}$ drastically decreases with $q$ since the decrease of $10^{\phi_p^{315\rm{K}}(6.5/q^3-5.5)}$ overcompensates $q^2$. However, this occurs up to $q$ around 2.5, which depends on $\phi_p^{315\rm{K}}$. In the limit of a very crosslinked microgel, one would get a slower release at temperatures below $T_c$, which is a consequence of having $D^{295\rm{K}}(\phi_p)<D^{315\rm{K}}(\phi_p)$ for all $\phi_p$. Notwithstanding, for reasonable values of $q$, above 1.1, one always gets $\tau_{1/2}^{295\rm{K}}/\tau_{1/2}^{315\rm{K}} < 1$. Finally, for $q>$3, we get that $\tau_{1/2}^{295\rm{K}}/\tau_{1/2}^{315\rm{K}}$ strongly depends on $\phi_p^{315\rm{K}}$ since it controls the reference state but $\tau_{1/2}^{295\rm{K}}$ yields practically the same values.

\section{Conclusions}

It is possible to scale up results from molecular dynamics to micro-sized systems by the DDFT theory to predict cargo and release halftimes of the order of seconds and more. In particular, for the release of phenol from homogeneous pNIPAM microgels with 23 $\mu$m of radius at 315 K, we capture the reported experimental~\cite{Si2019} halftime $\tau_{1/2}^{315\rm{K}} \simeq 2200$ s for $\phi_p^{315\rm{K}} \simeq$ 0.83. Additionally, this $\phi_p^{315\rm{K}}$ value is consistent with simulations~\cite{Kanduc2018} and relatively close to experimental data~\cite{Dong1990,Sasaki1999,vanDurme2004}. $\tau_{1/2}^{295\rm{K}}$ is given as a function of the degree of swelling of the hydrogel, $q$. We expect a similar $\tau_{1/2}$ behavior and even similar values for 5-FU than for phenol.

The DDFT needs the definition of $\phi_p$ as a function of the space, that is the particle shape, size, and composition, and the functions $\Delta G_{\rm{trans}}(\phi_p)$ and $D(\phi_p)$ as inputs. In addition, it needs boundary and initial conditions to completely define the problem. In our case, we get $\Delta G_{\rm{trans}}(\phi_p)$ for pNIPAM interacting with three types of cargo molecules, methane, phenol, and 5-FU. We use  Bennett's acceptance ratio method for the cargo transfer between bulk water and water-pNIPAM mixtures. Phenol and 5-FU produce similar  $\Delta G_{\rm{trans}}(\phi_p)$ trends, practically linear for all studied $\phi_p$ range, negative, and temperature independent (athermic). Conversely, the sign of $\Delta G_{\rm{trans}}(\phi_p)$ changes when crossing $T_c$ for methane from positive to negative  (endothermic). In all cases, $\Delta G_{\rm{trans}}$ keeps a linear behavior with $\phi_p$, contrasting with some theoretical predictions.

We used MD to compute function $D(\phi_p)$ for phenol in water-pNIPAM mixtures in the diluted limit at temperatures above and below $T_c$. We observe differences between the curves even when normalizing $D$ with the $D_0$ value for bulk water at the corresponding temperature. The one obtained above $T_c$ decays slower with $\phi_p$ than the one obtained below it, signaling that it is more difficult for a cargo molecule to transit across dispersed attractive chains than through aggregated ones at the same $\phi_p$.

In the case of phenol and for particles without cavity and small $\delta$ values, the release halftime can be approximated by $\tau_{1/2} \sim q^2 R_{\rm{ext}}^2/(6D)$, a process controlled by diffusion only. In this case, the free energy of transfer only rules the cargo amount to be transported by the microgel by fixing the partition coefficient. However, in general, diffusion and free energy affect the release halftime.

\section*{Conflicts of interest}

There are no conflicts to declare.

\section*{Acknowledgements}
The authors thank Prof. Matej Kanduč for helpful discussions. The authors thank CONACyT projects A1-S-9197 and A1-S-16002 for financial support. HAPR acknowledges a scolarship from CONACyT. A. M.-J acknowledges the financial support provided by the Spanish Junta de Andaluc\'{\i}a and European Regional Development Fund – Consejer\'{\i}a de Conocimiento, Investigaci\'{o}n y Universidad, Junta de Andaluc\'{\i}a (Projects PY20-00241 and A-FQM-90-UGR20).

\section*{Appendix: Calculation of the mean first passage time for a hollow spherical hydrogel}

In this section, we provide an analytical prediction of $\langle \tau \rangle$ for the release process across a spherical hollow microgel particle. For this purpose, we define the collapsed state as a reference and $q$ as the degree of swelling of the microgel. Hence, the radii of the shell of the hollow microgel are $qR_1$ and $qR_2$ at 295 K, where $R_1$ and $R_2$ are the corresponding radii in the collapsed state at 315 K. In addition, $q\delta$ is the halfwidth of the interface at 295 K and $\delta$ is its value at 315 K.
For a low concentration of cargo molecules, the cargo--cargo interactions are weak. Therefore, the release process can be modeled as a simple diffusion problem through a background effective potential induced by the hydrogel,
$u_\textmd{eff}(r)$~\cite{Klein1952,Ansari2000,Amato2017}. In other words, we
are interested in the time spent by the molecules
initially distributed homogeneosly inside the cavity to reach the external surface of the hydrogel, located at $r>q(R_2+2\delta)$ (we remind that $2\delta$ is the thickness of the shell interface). Direct integration of the Smoluchowski equation leads
to the following expression for the mean first passage time~\cite{Deutch1980}
\begin{equation}
\label{meanpassagetime}
\tau_{1/2} \simeq \langle \tau \rangle=\int_{s}^{q(R_2+2\delta)}dr\frac{e^{\beta
		u_\textmd{eff}(r)}}{D(r)r^2}\int_0^{r}dr^{\prime}{r^{\prime}}^2e^{
	-\beta u_\textmd{eff}(r^{\prime})} ,
\end{equation}
where the integral limit $s$ denotes the location of the center of mass of the molecules distributed inside the cavity at $t=0$, given by $s=q\sqrt{3/5}R_1$ for a cavity uniformly filled with cargo molecules.

In order to perform the double integral and calculate $\langle \tau \rangle$, we replace the smooth surfaces of the hydrogel shell with sharp interfaces, such that $u_\textmd{eff}(r)= u_\textmd{eff}$ for $qR_1<r<qR_2$, and $u_\textmd{eff}(r)=0$ elsewhere. Analogously, $D(r)= D$ for $qR_1<r<qR_2$ and $D(r)=D_0$ elsewhere. Inserting these two simple prescriptions into Eq.~\ref{meanpassagetime}, we find
\begin{eqnarray}
\label{meanpassagetime2}
\tau_{1/2} \simeq \langle \tau \rangle\!\!\!\!\!&=&\!\!\!\!\!\frac{q^2R_1^2}{15D_0}
+\frac{q^2}{6D} \Big(\frac { 2R_1^2(R_2\!-\!R_1) }{ R_2}(e^{
	\beta u_\textmd{eff}}-1)\!+\!R_2^2\!-\!R_1^2\Big) \nonumber  \\
&\!\!\!\!\!\!\!\!\!\!\!\!\!\!\!+&\!\!\!\!\!\!\!\!\!\!\!\!\!\frac{q^2}{6D_0}
\Big(\frac{ 4(R_2^3\!-\!R_1^3)\delta
}{ (R_2\!+\!2\delta)R_2 }(e^ {
	-\beta u_\textmd{eff}}\!-\!1)\!+\!4R_2\delta\! +\!4\delta^2 \Big).
\end{eqnarray}
As seen, the time scales as $\langle \tau \rangle\sim q^2$, as expected for a diffusive release. For release processes across a very high energy barrier, $\beta u_\textmd{eff}\gg 1$,$\tau_{1/2}$ is controlled by the exponential in the second
term, so $\langle \tau \rangle \sim q^2e^{\beta u_\textmd{eff}}/D$. Conversely, for strong cargo--hydrogel attractions, $\beta u_\textmd{eff}\ll -1$, the exponential in
the third term plays the key role, leading to $\langle \tau \rangle \sim q^2  e^{|\beta u_\textmd{eff}|}/D_0$. In addition, if the diffusion of the cargo is very slow inside the hydrogel shell compared to bulk diffusion, we find the scaling $\langle \tau \rangle \sim
q^2/D$.



\balance



\providecommand*{\mcitethebibliography}{\thebibliography}
\csname @ifundefined\endcsname{endmcitethebibliography}
{\let\endmcitethebibliography\endthebibliography}{}

\end{document}